\documentclass[twocolumn,showpacs,preprintnumbers,amsmath,amssymb]{revtex4}

\makeatletter

\usepackage{graphicx}
\usepackage{dcolumn}
\usepackage{bm}

\begin{document}

\title{Quantum wire fracture and discrete-scale invariance}

\author{Maiko Kikuchi\footnote{Present address:
Department of Materials Science and Engineering, Tokyo 
Institute of Technology 4259 Nagatsuda, Midori-ku, 
Yokohama 226-8502, Japan} and Masanori Yamanaka}
\affiliation{Department of Physics, College of Science and Technology,
Nihon University, Kanda-Surugadai 1-8-14, Chiyoda-ku,
Tokyo 101-8308, Japan}

\date{\today}

\begin{abstract}
This paper is a study of the behavior of experimentally observed
stress-strain force during the fracture of a quantum wire.
The magnitude of the force oscillates as a function of time 
and can be phenomenologically regarded as a sign 
of discrete-scale invariance. 
In the theory of discrete-scale invariance, 
termination of the wire is regarded as a phase transition. 
We estimate the critical point and exponents.
\end{abstract}


\maketitle

\section{introduction}

Discrete-scale invariance is ubiquitously found 
in catastrophic phenomena.~\cite{REF:sornette}
Typical examples are diffusion-limited-aggregation 
clusters~\cite{REF:DLA},
ruptures in heterogeneous systems~\cite{REF:rupture},
earthquakes~\cite{REF:eqrthq},
and financial crashes~\cite{REF:finance}.
The invariance is obtained by placing a restriction 
on the scale invariance, which requires complex critical exponents 
and log-periodic corrections to scaling. 
The corrections lead to oscillation in the observables 
and the periodicity of the oscillation becomes shorter 
as it approaches the critical point. 
Phenomenologically, this is regarded as the typical 
and universal property of discrete-scale invariance.

The strain force during the fracture process of a quantum wire,
in the first experiment of this type, have measured 
experimentally.~\cite{REF:takayanagi}
The magnitude of the force oscillates as a function of time, 
i.e., as a function of the external strain force.

In this paper we analyze the fracture process 
using discrete-scale invariance theory. 
Although discrete-scale invariance has in the past 
been studied in macroscopic systems, 
here we apply it to a microscopic system. 
We assume that the oscillation observed in the strain force 
intrinsically derives from the invariance. 
We estimate the critical exponents 
and make some conjectures on the results.

\section{singularities and log-periodic corrections}

In critical phenomena, the observables obey 
the Power Law near the critical point. 
This is expressed as
\begin{eqnarray}
f(x)\varpropto(x_c-x)^m,    
\label{eq:exponent1}
\end{eqnarray}
where $f(x)$ is the observable, $x$ is a parameter,
such as temperature, pressure, and so on, 
$x_c$ is the critical point, and $m$ is the critical exponent.
The Power Law reflects the scale invariance or self-similarity
of the underlying physics.
The exponent reflects the dimensionality and symmetry
of the system and is used to distinguish the universality class.
Discrete scale invariance theory states that critical phenomena
can have more general properties than the simple Power Law.
There is a complex critical exponent
\begin{eqnarray}
m=m'+m''i
\label{eq:exponent2}
\end{eqnarray}
where $m'$ and $m''$ are real numbers, and $i$ 
is the imaginary number unit. 
Putting (\ref{eq:exponent2}) to (\ref{eq:exponent1}), 
we have
\begin{equation}
  \begin{split} 
    f(x)& \varpropto Re[(x_c-x)^{m'+m''i}]\\
        & = Re[(x_c-x)^{m'}e^{im''\log(x_c-x)}]\\
        & = (x_c-x)^{m'}\cos\{ m''\log(x_c-x) \}\\
        & = (x_c-x)^{m'}[a_0 + \sum\limits_{n>0}\cos\{nd\log(x_c-x)+e\} ]
  \end{split}
\label{eq:exponent3}
\end{equation}
where $Re[\ \ ]$ denotes the real part 
and $d$ is a constant which is related 
to the preferred scaling ratio~\cite{REF:sornette}. 
By neglecting the higher-order terms in the Fourier series,
we obtain 
\begin{eqnarray}
f(x)=a+b(x_c-x)^{m'}[1+c \cos\{d\log(x_c-x)+e\}]
\label{eq:exponent4}
\end{eqnarray}
where $a,b$ and $c$ are constants.
This expresses the log-periodic oscillation 
superposed on the Power Law.

\section{Application to the experimental data}

A Power Law distribution of alternation detected 
in the strain force in a nanowire appears to be a signature 
of scale invariance, leading to the idea that a rupture 
in the nanowire can be regarded as a kind of "critical point."
In an analogy of the critical point, the rupture of a nanowire 
can be viewed as a cooperative phenomenon corresponding 
to the progressive buildup of stress and damage correlations. 
The rupture interaction increases exponentially on approaching 
the critical point, which may emerge as detectable signals 
exhibiting log-periodic oscillation patterns.

\begin{figure}[t]
\includegraphics[width=85mm]{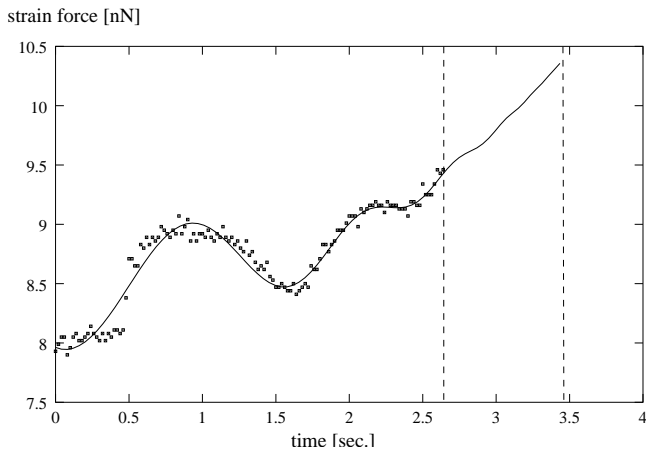}
\caption{
Strain force of the quantum wire as a funciton of time. 
Experiments~\cite{REF:takayanagi} are shown 
by the dots.
Fitting (\ref{eq:exponent6}) is shown by a solid line.
}
\label{fig:data}
\end{figure}

Therefore, we assume that the strain-stress, $F(t)$, 
can be described by eq. (\ref{eq:exponent4}) phenomenologically. 
It can be fitted using a modified version,
\begin{eqnarray}
F(t)=a+b(t_c-t)^{m'}[1+c \cos\{d \log(t_c-t)+e\}]+ft.
\label{eq:exponent5}
\end{eqnarray}
The last term is added to express the linear trend 
of subsidence.
Eq. (\ref{eq:exponent5}) contains 8 unknown parameters, 
$a,b,c,d,e,f,m',t_c$, which can be determined 
by the non-linear least-squares method. 
We estimated each of them to be
$a=5.48, b=0.282, c=0.252, d=-10.1, e=9.18, f=1.42, 
m'=1.98, t_c=3.46$, 
i.e. we obtain the phenomenological function,
\begin{multline}
F(t)=5.48+0.282(3.46-t)^{1.98} \\ 
\times[1+0.252\cos\{-10.1\log(3.46-t)+9.18\}]+1.42t,
\label{eq:exponent6}
\end{multline}
and show it in Fig.\ref{fig:data} as a solid curve.

\noindent

\section{Discussion}

Assuming discrete-scale invariance, 
i.e., based on log-periodic oscillation, 
we obtain the phenomenological function 
and estimate the critical point and index. 
There appears to be no explicit inconsistency 
between the experimental data and our fitting function 
at this phenomenological stage. 
Here we discuss the validity of the assumption.

As shown in Fig.\ref{fig:data}, 
the experimentally-observed rupture time is 2.60 seconds. 
In contrast, the estimated time by fitting is 3.46 seconds. 
Some possible reasons for this discrepancy are: 
(a) In the experiment, the nanowire ruptured 
before it reached the true critical point. 
(b) The experimental data does not have 
sufficient resolution. 

For (a), any external noise, 
such as a small shock to the sample or thermal fluctuation, 
may force an earlier termination of the wire.
If the experiment were performed under ideal conditions, 
i.e., under adiabatic stretching of the nanowire, 
and if the atoms were infinitely small, the rupture 
would be expected to occur at the true critical point. 
If we find any discrepancy, even in the adiabatic process, 
it is due to the finite volume effect of the atoms, 
since atoms cannot be subdivided on this energy scale. 
In this case, the experimentally-measured strength 
of the force is identical to that between 
single-atomic contact. 
For (b), the eq.(\ref{eq:exponent4}) has small 
and rapid oscillation in the limit $x \to x_c$. 
The amplitude becomes smaller and the period becomes shorter 
as we approach the critical point. 
If the amplitude of the oscillation is smaller 
than that of the resolution in the experiment, 
we cannot estimate true the critical point by fitting.

In this study, we assumed that discrete-scale invariance 
is applicable to microscopic systems. 
In this microscopic system, 
there are no explicit heterogeneous structures 
from the viewpoint of classical mechanics.
However, from a quantum mechanical point of view, 
the bonding networks of the wave function of electronic state 
of atoms may have a heterogeneous structure 
and would be expected to be reorganized 
as rupture approached while self-optimizing 
the total energy of the system.

Repeated experiments are desirable to confirm the validity 
of discrete-scale invariance 
and to distinguish quantum fracture from classical fracture. 
If the exponent takes a universal value, 
it would further support the assumption of invariance.

\section{acknowledgement}

We are grateful to 
S.~Miwa, S.~Shibata, Y.~Tanishiro, and K.~Takayanagi 
for allowing us to use their experimental data 
before publication, and to
S.~Miwa, C.~Akahori, and Y.~Tanishiro for valuable discussions.

\end{document}